\documentclass[12pt]{article}
\usepackage{latexsym}
\thicklines                         % thick straight lines for pictures (LaTeX)
\def\footnoteitem(#1)#2{
\begin{list}{#1}{\labelwidth4.0mm \leftmargin7.0mm
\labelsep2.5mm \rightmargin7.0mm \parsep0.5ex plus0.2ex minus0.1ex
\itemsep0ex plus0.2ex }
\item #2
\end{list}
}

\newcommand{\be}{\begin{equation}}
\newcommand{\ee}{\end{equation}}
\newcommand{\ba}{\begin{eqnarray}}
\newcommand{\ea}{\end{eqnarray}}

\newcommand{\cD}{{\cal D}}

\newcommand{\tr}{{\rm tr}}
\newcommand{\Det}{{\rm det}}
\newcommand{\half}{\frac{1}{2}}
\newcommand{\cDH}{{\cD_{\rm H}}}
\newcommand{\cbar}{{\overline{c}}}
\newcommand{\Ap}{{A'}}
\newcommand{\thetap}{{\theta'}}

\begin{document}
%%%%%%%%%%%%%%%%%%%%%%%%%%%%%%%%%%%%%%%%%%%%%%%%%%%%%%%%%%%%%%%%%
%
\begin{titlepage}

\vskip 3mm
\rightline{UT~CCP-P-83}
\rightline{March 2000}

\baselineskip=20pt plus 1pt
\vskip 0.5cm

\centerline{\LARGE Nonperturbative Gauge Fixing and Perturbation Theory}
\vskip 1.0cm
\centerline{\large Wolfgang Bock$^{a,1}$, Maarten Golterman$^{b,c,2}$,}
\centerline{\large Michael Ogilvie$^{c,3}$ and Yigal Shamir$^{d,4}$}
\vskip 1.0cm
\centerline{\em {}$^a$Institute of Physics, University of Siegen,
57068 Siegen, Germany}
\centerline{\em  {}$^b$Center for Computational Physics, University
of Tsukuba, Tsukuba,}
\centerline{\em Ibaraki 305-8577, Japan}
\centerline{\em  {}$^c$Department of Physics, Washington University,
St. Louis, MO 63130, USA}
\centerline{\em {}$^d$Beverly and Raymond Sackler Faculty of Exact Sciences,
Tel-Aviv University,}
\centerline{\em Ramat Aviv 69978, Israel}
\vskip 1.0cm
\baselineskip=12pt plus 1pt
\parindent 20pt
\centerline{\bf Abstract}
\textwidth=6.0truecm
\medskip

\frenchspacing
\noindent
We compare the gauge-fixing approach
proposed by Jona-Lasinio and Parrinello, and by Zwanziger (JPLZ)
with the standard Fadeev-Popov procedure,
and demonstrate perturbative equality of gauge-invariant
quantities, up to irrelevant terms induced by the cutoff.
We also show how a set of local,
renormalizable Feynman rules can be constructed for the
JPLZ procedure.

\nonfrenchspacing

\vskip 0.8cm
\vfill
%\begin{flushleft}
\noindent $^1$ e-mail: {\em bock@physik.uni-siegen.de}  \\
\noindent $^2$ e-mail: {\em maarten@aapje.wustl.edu } \\
\noindent $^3$ e-mail: {\em mco@morgan.wustl.edu } \\
\noindent $^4$ e-mail: {\em shamir@post.tau.ac.il}  \\
\end{titlepage}

{\it 1.}
Gauge fixing in Yang--Mills theories is well understood in perturbation
theory.  But outside of perturbation theory the situation is
more complicated, because of the existence of Gribov copies \cite{gribov}.  
These complications arise as soon as the theory
is nonlinear, which is the case for nonabelian gauge theories, but
also, for instance, with a nonlinear gauge-fixing condition 
in an abelian theory.

This does not mean that Yang--Mills theories do not exist outside
perturbation theory.  It is well known that they can be defined on
the lattice, using the compactness of the group to dispense with
gauge fixing altogether.  However, nonperturbative gauge fixing
is interesting in a variety of contexts, ranging from the
need to make contact between lattice and continuum calculations
(see for example Ref.~\cite{mandula})
to the definition of chiral gauge theories on the lattice
(see for example Ref.~\cite{bgls}).  For a discussion of the
numerical implementation of the gauge-fixing method discussed in
this paper, see Ref.~\cite{hentyetal}.

One can try to extend the standard BRST construction to the
lattice.  When one does this, one finds, quite generally and rigorously,
that the BRST gauge-fixed partition function vanishes
\cite{neuberger}.  This result can be
heuristically explained 
as the result of pairwise cancellations of
lattice Gribov copies, which
occur with opposite signs of the Faddeev--Popov determinant.
(For some recent work on overcoming this problem, 
see Refs.~\cite{schaden,testa}.)

A different nonperturbative method for gauge fixing has been proposed
some time ago by Jona-Lasinio and Parrinello and by Zwanziger (JLPZ)
\cite{jlp,zwanziger} (see also Ref.~\cite{kerler}).  
Starting from the euclidean ungauged partition function
\be
Z=\int\cDH A\;{\rm exp}\left(-S_{\rm inv}(A)\right)\,,
\label{ZUG}
\ee
one inserts one in the form
\be
1=\frac{\int\cDH h\;{\rm exp}\left(-S_{\rm ni}(A^h)\right)}
{\int\cDH h\;{\rm exp}\left(-S_{\rm ni}(A^h)\right)}
\label{ONE}
\ee
into the functional integral.
Here $\cDH A$ and $\cDH h$ denote the invariant measures over
the gauge field $A$ and the group-valued scalar field $h$,
respectively; $A^h$ is the gauge transform of $A$ under a
gauge transformation $h$.  $S_{\rm ni}(A)$ is any gauge
noninvariant local functional of $A$.
Because of the invariance of
$\cDH A$ and $S_{\rm inv}(A)$, we can perform a gauge
transformation in the numerator of Eq.~(\ref{ZUG}) with
Eq.~(\ref{ONE}) inserted, and, dropping a trivial factor
$\int\cDH h$, we obtain
\be
Z=\int\cDH A\;\frac{{\rm exp}\left(-S_{\rm inv}(A)-S_{\rm ni}(A)
\right)}
{\int\cDH h\;{\rm exp}\left(-S_{\rm ni}(A^h)\right)}\,.
\label{ZGF}
\ee
This procedure is completely rigorous on the lattice,
and sources coupled to gauge-invariant operators can be
added without changing the argument.  We also note that
the Boltzmann weights for both integrations contained in
Eq.~(\ref{ZGF}) are positive. 

While this method is conceptually very simple, 
much less is known about how this nonperturbative
method works in perturbation theory, as compared with
the standard gauge fixing based on BRST invariance.
Note, however, that there are 
some similarities between the two procedures.
A noninvariant term $S_{\rm ni}(A)$ is added to the action,
and (for a suitable choice)
this will make the quadratic part of the action invertible.
The contributions of different orbits are weighted properly
because of the group integral in the denominator of Eq.~(\ref{ZGF}),
which plays a role 
similar to that of the Faddeev--Popov (FP) determinant in the usual
case.  In the FP case, however, the determinant can be expressed
as an integral over ghosts, and the gauge-fixed action including
the ghost terms is local, whereas here this is not the case.
This is an important difference, because 
locality is a key ingredient in power-counting arguments, and
thus at the heart of the usual perturbative analysis of
renormalization.

It is therefore of interest to find out whether perturbation theory
can be systematically developed for a JLPZ gauge-fixed Yang--Mills
theory, and its relation to the usual FP results.  
This question has been addressed previously by Fachin
\cite{fachin}, who analyzed the vacuum polarization for the
choice $S_{\rm ni}(A)= \tr(M^2 A_\mu^2)$ at one loop.
He concluded that for $M\to\infty$ the transverse
part of the vacuum polarization agrees with that
obtained using the FP method; the longitudinal
part vanishes for $M\to\infty$, as in Landau gauge.  This
equivalence for $M\to\infty$ at fixed cutoff was already 
derived formally in Ref.~\cite{jlp}, and
another formal discussion appeared recently in Ref.~\cite{fujter}.
Here, we are interested in considering the situation where
$M$ is chosen to be of the same order of magnitude as the cutoff.
The transverse part of the one-loop vacuum polarization with JLPZ
gauge fixing, for example, was found to contain terms
proportional to $p^2\delta_{\mu\nu}-p_\mu p_\nu$ times
$p^2/M^2$ or $p^4/M^4$ \cite{fachin}, so, if
we choose $M\sim\Lambda$ (with $\Lambda$ the cutoff), 
such terms are of order $1/\Lambda^2$, and vanish
when we take $\Lambda\to\infty$.
We present below a general argument that, if we choose $M$ to
be fixed in units of the cutoff, perturbation theory for a
JLPZ gauge-fixed theory is equivalent to that of the same
theory gauge-fixed using the standard FP method.  
We discuss in some detail what ``equivalent" means
in this context.  We also show that our arguments for the
equivalence of JLPZ and FP gauge fixing can be used to construct
a set of local Feynman rules for the JLPZ gauge-fixed gauge theory.

\bigskip
{\it 2.} We begin with the JLPZ gauge-fixed partition function 
in the presence of
a source for the gauge field $A_\mu$ (working in
euclidean space-time):
\be
Z(J)=\int\cDH A\;\frac{{\rm exp}\left(-S_{\rm inv}(A)-S_{\rm ni}(A)
+J_\mu A_\mu\right)}
{\int\cDH h\;{\rm exp}\left(-S_{\rm ni}(A^h)\right)}\,,
\label{Z}
\ee
where $S_{\rm inv}(A)$ is the gauge-invariant classical action,
and $S_{\rm ni}(A)$ is not invariant.  
As noted before, $A^h$ is the gauge transform of $A$ under a finite
gauge transformation,
\be
A_\mu^h=h(A_\mu+\frac{i}{g}\partial_\mu)h^\dagger\;.\label{GT}
\ee
We will take
\ba
S_{\rm inv}(A)&=&\frac{1}{2}\tr(F_{\mu\nu}^2)\;,\label{ACTION} \\
S_{\rm ni}(A)&=&\tr(M^2A_\mu^2)\;,\nonumber
\ea
where we do not indicate the integration $\int d^4x$ explicitly, 
and $M$ a parameter with the dimension of a mass.  We will assume
that $M$ is proportional to the cutoff for the case
of regulators such as the lattice, in which case
$M$ is proportional to the inverse lattice spacing.
However, it is sufficient for our arguments that
$M$ is chosen large compared to all physical scales.
This assumption applies in the case of dimensional regularization.
Also, $A_\mu=A_\mu^a T^a$, with $T^a$ the generators 
of the gauge group, with 
\ba
\tr(T^aT^b)&=&\frac{1}{2}\delta^{ab}\;,\label{NORM} \\
{}[T^a,T^b]&=&if^{abc}T^c\;.\label{STRCSTS}
\ea
If we now consider perturbation theory, and use a
regulator without power-like divergences
such as dimensional regularization, we may extend the range
of integration of the variables $A_\mu^a$ from $-\infty$ to
$\infty$. The invariant measure is just the
flat measure
\be
\cDH A=\cD A=\prod_{x,\mu,a}dA_\mu^a(x)\;, \label{FLAT}
\ee
and similarly for $\cDH h$ (after parametrizing $h$ as in
Eq.~(\ref{HEXP}) below).  

We wish to show that this gauge-fixed partition function is
equivalent (in a way to be discussed in more detail in the
next section), under certain
assumptions which are valid in perturbation theory, 
to the standard FP gauge-fixed form for the
covariant gauge $\partial_\mu A_\mu=0$.  The derivation 
is based on two ingredients.  The first ingredient is that
we insert a constant into the partition 
function, written in the form
\be
\mbox{constant}=
\Det(\Box)\int\cD\eta\;\delta(\partial_\mu A_\mu-\frac{1}{M}
\Box\eta)\;,\label{ETAINT}
\ee
where $\cD\eta$ denotes the (flat) measure for $\eta$, 
a new field which takes values in the Lie algebra of the gauge group.
We may choose boundary conditions such as to avoid the trivial
zero mode of $\Box$.
The second ingredient is to make use of the fact that 
the physics at scales below $M$ is not altered by
adding or changing terms which are irrelevant
in the sense of Wilson's renormalization group.

First, let us expand $S_{\rm ni}(A^h)$ in $g/M$, writing
\be
h={\rm exp}(i\frac{g}{M}\theta)\;.\label{HEXP}
\ee
We find
\be 
S_{\rm ni}(A^h)=\half M^2(A_\mu^a)^2-M\theta^a\partial_\mu A_\mu^a
+\half\partial_\mu\theta^a (D_\mu(A)\theta)^a+
O\left(\frac{g}{M}\right)\;,\label{SNIEXP}
\ee
where
\be
(D_\mu(A)\theta)^a=\partial_\mu\theta^a+gf^{abc}A_\mu^b\theta^c\;.
\label{COVDER}
\ee
Note that $-\partial_\mu D_\mu(A)$ is just the FP
operator for the covariant gauge.

We then shift the vector field 
\be
A_\mu\to \Ap_\mu=A_\mu- \frac{1}{M}\partial_\mu\eta\;. 
\label{SHIFT}  
\ee
This gives
\ba
S_{\rm ni}(A^h)&=&\half M^2(\Ap_\mu^a)^2+M\Ap_\mu^a\partial_\mu\eta^a
+\half(\partial_\mu\eta^a)^2 \label{DENOM} \\
&&-M\theta^a\partial_\mu \Ap_\mu^a-\theta^a\Box\eta^a
+\half\partial_\mu\theta^a
(D_\mu(\Ap)\theta)^a+O\left(\frac{g}{M}\right)\;,\nonumber
\ea
and
\be
S_{\rm inv}(A)+S_{\rm ni}(A)=S_{\rm inv}(\Ap)+
\half M^2(\Ap_\mu^a)^2+M\Ap_\mu^a\partial_\mu\eta^a
+\half(\partial_\mu\eta^a)^2+O\left(\frac{g}{M}\right)\;.\label{NUM}
\ee
The terms $\half M^2(\Ap_\mu^a)^2+M\Ap_\mu^a\partial_\mu\eta^a
+\half(\partial_\mu\eta^a)^2$ cancel between the numerator and
denominator of the integrand in Eq.~(\ref{Z}).  
The terms $M\theta^a\partial_\mu \Ap_\mu^a$ and 
$M\Ap_\mu^a\partial_\mu\eta^a$ (which is equal to
$-M\eta^a\partial_\mu\Ap_\mu^a$ by partial integration)
may be dropped because of the $\delta$-function,
$\delta(\partial_\mu\Ap_\mu)$, in Eq.~(\ref{ETAINT}).   

All the $O(g/M)$ terms are irrelevant, and may
therefore be omitted without changing the (renormalized) theory.
Doing this, we can perform the $h$-integral in the denominator
of the integrand in Eq.~(\ref{Z}), obtaining
\ba
Z(J)&=&\Det(\Box)\int\cD\eta\cD \Ap\;\delta(\partial_\mu \Ap_\mu)\;
\Det^{1/2}(-\partial_\mu D_\mu(\Ap)) \label{HALFWAY} \\
&&\hspace{1.5truecm}\times{\rm exp}\left(-S_{\rm inv}(\Ap)
+J_\mu A_\mu-\half\eta\Box
(-\partial_\mu D_\mu(\Ap))^{-1}\Box\eta\right)\;.\nonumber
\ea
Using the fact that physical quantities do not change when
we replace the source term $J_\mu A_\mu$ by a new source term
$J'_\mu \Ap_\mu$ coupling to $\Ap_\mu$ instead of to
$A_\mu$ ($A_\mu$ and $\Ap_\mu$ 
are equivalent interpolating fields), we 
can now also perform the $\eta$ integral. Dropping the primes,
we obtain
\be
Z(J)=\int\cD A\;\delta(\partial_\mu A_\mu)\;\Det(-\partial_\mu D_\mu(A))
\;{\rm exp}\left(-S_{\rm inv}(A)+J_\mu A_\mu\right)\;.\label{ALMOST}
\ee
Representing the $\delta$-function as
\be
\delta(\partial_\mu A_\mu^a)\propto
\lim_{\xi\to 0}\,{\rm exp}\left(-\frac{1}{2\xi}(\partial_\mu
A_\mu^a)^2\right)\;, \label{DF}
\ee
and introducing an algebraic field $B$ and ghost fields
$c$ and $\cbar$, this can be recast as
\ba
&&Z(J)=\lim_{\xi\to 0}\int\cD A\cD B\cD\cbar\cD c \label{FINAL}\\
&&\hspace{3truecm}\times
{\rm exp}\left(-S_{\rm inv}(A)-\frac{1}{2}\xi B^2+iB\partial_\mu A_\mu
-\cbar(-\partial_\mu D_\mu(A))c+J_\mu A_\mu\right). \nonumber
\ea
This is the standard BRST-invariant
form of the FP gauge-fixed partition function in Landau gauge. 
Of course, correlation functions of gauge-invariant operators
are independent of $\xi$.  We observe that, if we would
take the limit $M\to\infty$ {\it before} removing the cutoff,
our argument constitutes an alternative derivation of the
equivalence of Eq.~(\ref{Z}) to Landau gauge given in Ref.~\cite{jlp}.
Note that the standard way of gauge fixing employed in
lattice QCD computations is formally equivalent to this limit.
What is new here, is that we do not take the limit $M\to\infty$
first, but keep it at the order of the cutoff.
Nevertheless, the parameter
$M$ has disappeared from Eq.~(\ref{FINAL}).

The derivation given above is valid only in perturbation theory. 
We assumed that the $\theta$-
and $\eta$-integrals converge, {\it i.e.} that 
the FP operator $-\partial_\mu D_\mu(A)$ has only 
positive eigenvalues.  This is not in general the case, but it
is true in perturbation theory.  

If we use a regulator with a hard cutoff, such as the lattice, additional
subtractions will be needed in order to remove power-like divergences,
which may appear as a consequence of dropping irrelevant terms.
Also note that, even though Eq.~(\ref{SHIFT}) is linear, in general
the invariant measure written in terms of the
Lie-algebra valued fields is nonlinear for such a regulator, 
and it would therefore
change under this transformation.  However, this nonlinearity is
proportional to the coupling constant $g$, and therefore the
effects of this shift are of order $g/M$.  

\bigskip
{\it 3.} In this section, we will 
discuss in more detail what we mean by ``equivalent."
It is clear that, in general, correlation functions 
of the form $\langle A^a_{\mu}(x)A^b_\nu(y)\dots\rangle$
are not the same in the JLPZ and FP versions.  A trivial  
example is the $O(g^0)$ two-point
function $\langle A_\mu(p)A_\nu(q)\rangle=\delta(p+q)
G_{\mu\nu}(p)$, with $G_{\mu\nu}(p)$ equal to
\ba
G^{\rm JLPZ}_{\mu\nu}(p)&=&
\frac{1}{p^2}\left(\delta_{\mu\nu}-
\frac{p_\mu p_\nu}{p^2}\right)+\frac{1}{M^2}
\frac{p_\mu p_\nu}{p^2}\;, \label{TWOP} \\
G^{\rm FP}_{\mu\nu}(p)&=&
\frac{1}{p^2}\left(\delta_{\mu\nu}-
\frac{p_\mu p_\nu}{p^2}\right)+\frac{\xi}{p^2}
\frac{p_\mu p_\nu}{p^2}\;,\ \ \ \xi\to 0\;, \nonumber
\ea
in the theories defined by Eqs.~(\ref{Z}) and (\ref{FINAL}), 
respectively.  This is not in contradiction with the general
argument presented above, because of the change of interpolating
field, $A_\mu\to\Ap_\mu$. 

The equality of the transverse part of the two-point function
at tree level (Eq.~(\ref{TWOP})) suggests in what sense the
two theories defined by Eqs.~(\ref{Z}) and (\ref{FINAL}) are
equivalent.  These two theories are not identical,
because of the fact that we dropped irrelevant terms in going
from the JLPZ to the FP version.  The correct statement is that
physical quantities (which are necessarily extracted from
correlation functions of gauge-invariant operators) in the JLPZ
version can be mapped into those of the FP version by a 
finite renormalization of the bare coupling constant $g$.
In the FP version of the theory the fact that only a coupling
constant renormalization is needed follows from BRST invariance.
In the JLPZ version, the same follows from the observation
that JLPZ gauge fixing can be ``undone" by multiplying
Eq.~(\ref{Z}) (for $J_\mu=0$) by a constant in the form 
$\int \cDH g$, and 
transforming $A_\mu\to A_\mu^g$, removing $S_{\rm ni}$ from
the partition function.  
Wave-function renormalizations are not necessary for physical 
quantities.

In other words, renormalized perturbation theory for physical
quantities is the same in both versions for a suitable
definition of the renormalized coupling constant, but the 
relation between the renormalized and bare coupling constants
is, in general, different in the FP and JLPZ versions of the theory.

\bigskip
{\it 4.} In the case of noninvariant correlation functions,
it is well known that,
in the FP version of the theory, only a universal
coupling-constant renormalization and multiplicative
wave-function renormalizations are necessary as a consequence
of BRST invariance.  The situation is less clear in the JLPZ
version of the theory.
Some progress can be made however, by observing that a set of
local and renormalizable Feynman rules 
can be constructed for the JLPZ partition function, Eq.~(\ref{Z}).
Start with diagonalizing the quadratic form in $\theta$ and $\eta$ in
Eq.~(\ref{DENOM}) by a shift $\theta=\theta'-\eta$.
After this shift, Eq.~(\ref{DENOM}) can be written as the sum
of two parts,
\be
S_{\rm ni}(A^h)=S^\eta_{\rm ni}(\Ap,\eta)+S^\theta_{\rm ni}(\Ap,\eta,
\thetap)\;, \label{SPLIT}
\ee
with $S^\theta_{\rm ni}(\Ap,\eta,\thetap=0)=0$.  To order $g/M$
we obtain
\ba
S^\eta_{\rm ni}(\Ap,\eta)&=&\half M^2(\Ap_\mu^a)^2+
\frac{1}{2}gf^{abc}\partial_\mu \eta^a A_\mu^b
\eta^c+O\left(\frac{g}{M}\right)\;, \label{SETA} \\
S^\theta_{\rm ni}(\Ap,\eta,\thetap)&=&\half\partial_\mu\thetap^a
(D_\mu(\Ap)\thetap)^a-gf^{abc}\partial_\mu\thetap^a\Ap_\mu^b\eta^c+
O\left(\frac{g}{M}\right)\nonumber
\ea
upon using $\partial_\mu\Ap_\mu=0$.
Next, we define the following
action (dropping the primes on $A_\mu$ and $\theta$):
\ba
S^{\rm JLPZ}_{\rm pert}&=&
S_{\rm inv}(A+\partial\eta/M)+S_{\rm ni}(A+\partial\eta/M)
-S^\eta_{\rm ni}(A,\eta)
+S^\theta_{\rm ni}(A,\eta,\theta) \label{JLPZPERT} \\
&=&\frac{1}{4}(F_{\mu\nu}^a)^2+\frac{1}{2}(\partial_\mu\eta^a)^2
-\frac{1}{2}gf^{abc}\partial_\mu \eta^a A_\mu^b
\eta^c \nonumber \\
&&
+\half\partial_\mu\theta^a (D_\mu(A)\theta)^a
-gf^{abc}\partial_\mu\theta^a A_\mu^b\eta^c
+O\left(\frac{g}{M}\right)\;. \nonumber
\ea
Here ``$O(g/M)$" indicates all higher order terms in the expansion
in $g$, including those coming from the shift Eq.~(\ref{SHIFT}), 
and they should be kept (up to the order in $g$ of interest).
This action will give rise to a correct set of Feynman rules
if we add the rule that a factor $-1$ be applied for each
connected $\theta$-subdiagram without external $\theta$ lines.
This is reminiscent of
the minus sign for ghost loops in the FP case:  the
integral over $\theta$ in the denominator of Eq.~(\ref{Z}) gives
rise to an effective action $S_{\rm eff}(A,\eta)$,
\be
{\rm exp}(S_{\rm eff}(A,\eta))=
\int\cD\theta\;{\rm exp}\left(-S^\theta_{\rm ni}(A,\eta,\theta)\right)
\;, \label{SEFF}
\ee
the vertices of which correspond precisely to these connected
$\theta$-subdiagrams.
The additional minus sign corresponds to the fact that this
integral appears in the denominator of the integrand in Eq.~(\ref{Z}).
If we work with a regulator
in which it is important to keep the nonlinear terms in the
measure, this can be taken into account by treating these nonlinear
terms as part of the action.

Finally, there is still the $\delta$-function of Eq.~(\ref{ETAINT}),
through which the field $\eta$ was introduced.  After the shift
Eq.~(\ref{SHIFT}), this is just $\delta(\partial_\mu A_\mu)$,
which again can be represented as in Eq.~(\ref{DF}).
This leads to a set of Feynman rules for a local theory,
Eq.~(\ref{JLPZPERT}), which is
renormalizable by power counting 
for any value of $\xi$, with the original JLPZ version (Eq.~(\ref{Z}))
corresponding to the limit $\xi\to 0$ (at fixed cutoff).

It is instructive to compare our local Feynman rules with the non-local
Feynman rules of Ref.~\cite{fachin}.  Our Feynman rules are defined
by Eq.~(\ref{JLPZPERT}), with the minus-sign rule for connected
$\theta$-subdiagrams, and the $\delta$-function constraint on $A_\mu$,
enforced by taking $\xi\to 0$ in Eq.~(\ref{DF}).  Note that the
longitudinal part of the gauge field is represented by $\eta$,
through Eq.~(\ref{ETAINT}).  The Feynman rules of Ref.~\cite{fachin}
were derived by integrating out the field $\theta$, or equivalently
$h$, in Eq.~(\ref{Z}), and without introducing the field $\eta$
to represent the longitudinal part of the gauge field.  This leads
to non-local vertices, where the non-locality arises from the
integration over $\theta$, as well as from the non-local relation between
$\eta$ and the longitudinal part of $A_\mu$.

It is straightforward to obtain these non-local Feynman rules
from our Feynman rules by (perturbatively) integrating out $\theta$
in Eq.~(\ref{JLPZPERT}), and by converting the Feynman rules for $\eta$
to Feynman rules for the longitudinal part $A_{L\mu}$ of $A_\mu$.  This
is obvious, because both our Feynman rules and those
of Ref.~\cite{fachin} are derived from the same partition function,
Eq.~(\ref{Z}).  In momentum space, the relation between $\eta$ and
the longitudinal part $A_{L\mu}$ of $A_\mu$ is
\begin{equation}
\eta(p)=-iM\frac{p_\mu}{p^2}A_{L\mu}(p)\;,\label{ETAA}
\end{equation} 
and from $\langle\eta(p)\eta(q)\rangle=\delta(p+q)/p^2$ we thus find
\begin{equation}
\frac{p_\mu p_\nu}{p^2}\langle A_{L\mu}(p)A_{L\nu}(q)\rangle=
\frac{1}{M^2}\delta(p+q)\;,\label{PROPS}
\end{equation}
in accordance with the first equation in Eq.~(\ref{TWOP}), as well as 
with Ref.~\cite{fachin}.
The simplest non-local vertex of Ref.~\cite{fachin} corresponds
to the $\partial\eta A\,\eta$ three-point vertex of Eq.~(\ref{JLPZPERT}),
which reads in momentum space, with $k$ the momentum of $A_\mu$ and
$p,q$ the $\eta$ momenta,
\begin{equation}
-\frac{1}{2}gf^{abc}i(p_\mu-q_\mu)\;. \label{ETAVERTEX}
\end{equation}
Using Eq.~(\ref{ETAA}), this translates into a
non-local three-point vertex for the gauge fields:
\begin{equation}
\frac{1}{6}M^2gf^{abc}\left[i(p_\mu-q_\mu)\frac{p_\kappa q_\lambda}{p^2q^2}
+i(q_\kappa-k_\kappa)\frac{q_\lambda k_\mu}{q^2k^2}
+i(k_\lambda-p_\lambda)\frac{k_\mu p_\kappa}{k^2p^2}\right]\;,\label{AVERTEX}
\end{equation}
where we symmetrized in the three gauge-field lines.  This is exactly the
non-local three-point gauge-field vertex of Ref.~\cite{fachin}.  
The other non-local vertices of the Feynman rules of Ref.~\cite{fachin}
can be obtained in a similar way.  

It follows that, for correlation functions
involving only gauge fields on the external lines, one obtains the
same result from either our Feynman rules or those of Ref.~\cite{fachin}.
This is true in particular for the one-loop vacuum polarization,
which was calculated explicitly in Ref.~\cite{fachin}.  The result found there
agrees with the vacuum polarization in Landau gauge, {\it i.e.} as
calculated from Eq.~(\ref{FINAL}).

We conclude this section by noting that other field redefinitions
of the field $\theta$ can be used, for instance
\be
{\rm exp}\left(i\frac{g}{M}\theta\right)=
{\rm exp}\left(i\frac{g}{M}\thetap\right)\;
{\rm exp}\left(-i\frac{g}{M}\eta\right)\;, \label{OTHER}
\ee
or
\be
{\rm exp}\left(i\frac{g}{M}\theta\right)=
{\rm exp}\left(i\frac{g}{M}\thetap\right)\;
\left(1-i\frac{g}{M}\eta\right)\;. \label{OTHERTOO}
\ee
These examples differ only by terms of order $g/M$ from the
one employed above, and thus lead to a different specific form
of the $O(g/M)$ terms in Eq.~(\ref{JLPZPERT}).

\bigskip
{\it 5.} We argued that the gauge-fixing procedure proposed in
Refs.~\cite{jlp,zwanziger}, with the choice of
Eq.~(\ref{ACTION}) for $S_{\rm ni}(A)$, is perturbatively
equivalent to the standard gauge-fixing procedure with Fadeev--Popov
ghosts.  In our derivation, we chose $M$ to be of the order
of the cutoff, thus
extending earlier arguments in which the (formal) limit $M\to\infty$
was considered.
 
``Equivalent" here means that perturbatively calculated
relations between physical quantities will be the same in both
versions of the theory.  Since, in addition, the JLPZ method leads to a
weighting of the integration over orbits which takes Gribov
copies correctly into account, this method may be the preferred
one for nonperturbative calculations in gauge-fixed Yang--Mills
theories.

We also derived local Feynman rules for the JLPZ version of the 
theory, Eq.~(\ref{Z}), from which it can be seen that the theory
is renormalizable by power counting. The choice of
$M$ at the order of the cutoff is a key ingredient here.
By construction, correlation
functions of gauge-invariant operators are the same when
calculated perturbatively from either Eq.~(\ref{Z}) or 
Eq.~(\ref{JLPZPERT}), after
the field redefinition of Eq.~(\ref{SHIFT})
is taken into account.
In order to renormalize correlation
functions of gauge noninvariant operators, counterterms may
have to be added to Eq.~(\ref{JLPZPERT}), in addition to those 
needed to renormalize gauge-invariant quantities.  The locality guarantees
that all counterterms necessary for renormalization are local, and
the renormalizability guarantees that only a finite number, all
with mass dimension less than or equal to four, will be needed.
It is also clear that this can be done in such a way that the
invariance of gauge-invariant correlation functions is maintained,
because of the gauge invariance of the original formulation of
Eq.~(\ref{Z}).
Hence, we believe that no problems will be encountered in carrying
out this program order by order in perturbation theory, for the
theory of  Eq.~(\ref{JLPZPERT}).  But it is not clear how this
would then ``translate back" to a JLPZ-like formulation as in
Eq.~(\ref{Z}).  What is
lacking is a tool similar to BRST symmetry in the FP version of 
the theory, which could be used to further 
control the form of the counterterms. 
It would be interesting and useful if such
a mechanism could be found.

We end with a comment on our use of the 
specific form of the action $S_{\rm ni}(A)$ in Eq.~(\ref{ACTION}). 
This choice is the ``most" (and only) relevant local operator in the 
sense of the renormalization
group.  Our analysis does not work when $S_{\rm ni}(A)$ is
chosen to be a marginal operator.  We expect that they will
work if we would add a marginal operator to $S_{\rm ni}(A)$
of Eq.~(\ref{ACTION}).  For instance, a term of the form
$c\sum_\mu\tr(A_\mu^4)$, with $c$ a constant of order $g^2$ 
(which is natural on the lattice), can
be removed by a field redefinition of the form $A_\mu\to
A_\mu+(c/2)A_\mu^3/M^2$.  Since the nonlinear term of this
field redefinition is of order $1/M^2$, this will just remove the
term $\sum_\mu\tr(A_\mu^4)$, without introducing any other 
marginal terms.

\bigskip
\leftline{\it Acknowledgements}
MG would like to thank Giancarlo Rossi, Massimo
Testa and Arjan van der Sijs
for useful discussions, and the Physics Department of
the University of Rome II ``Tor Vergata" for hospitality.  
MG and MO are supported in part
by the US Department of Energy, and YS is supported in part by
the Israel Academy of Science.

\end{document}